\documentclass{aa}
\usepackage{psfig}

\newcommand{\MC}{\multicolumn}
\newcounter{qub}
\newcommand{\qq}{\addtocounter{qub}{1}\arabic{qub}}

\begin{document}
\thesaurus{20(04.19.1; 11.06.2; 11.04.1; 11.19.3; 11.03.2)}
\title{ The Hamburg/SAO Survey for Emission--Line Galaxies }
\subtitle{ IV. The Fourth List of 119 Galaxies }

\author{%
A.Y.~Kniazev\inst{1}\fnmsep\inst{8}
\and D.~Engels\inst{2}
\and S.A.~Pustilnik\inst{1}\fnmsep\inst{8}
\and A.V.~Ugryumov\inst{1}\fnmsep\inst{8}
\and T.F.~Kniazeva\inst{1}
\and A.G.~Pramsky\inst{1}
\and N.~Brosch\inst{3}
\and H.-J.~Hagen\inst{2}
\and U.~Hopp\inst{4}
\and Y.I.~Izotov\inst{5}
\and V.A.~Lipovetsky\inst{1}\fnmsep\thanks{Deceased 1996 September 22.}
\and J.~Masegosa\inst{6}
\and I.~M\'arquez\inst{6}
\and J.-M.~Martin\inst{7}
}

\offprints{akn@sao.ru}

\institute{
Special Astrophysical Observatory, Nizhnij Arkhyz, Karachai-Circessia,
369167, Russia
\and Hamburger Sternwarte, Gojenbergsweg 112, D-21029 Hamburg, Germany
\and Wise Observatory, Tel-Aviv University, Tel-Aviv 69978, Israel
\and Universit\"atssternwarte M\"unchen, Scheiner Str. 1, D-81679 M\"unchen, Germany
\and Main Astronomical Observatory, Goloseevo, Kiev-127, 03680, Ukraine
\and Instituto de Astrofisica de Andalucia, CSIC, Aptdo. 3004, 18080, Granada, Spain
\and D\'epartement de Radioastronomie ARPEGES, Observatoire de Paris, F-92195 Meudon Cedex, France
\and Isaac Newton Institute of Chile, SAO Branch
}

\date{Received \hskip 2cm; Accepted}

\maketitle

\markboth{A.Kniazev et al.: The Hamburg/SAO Survey for Emission-Line
Galaxies. IV}
{The Hamburg/SAO Survey. IV}

\begin{abstract}

We present the fourth list with results\footnote{Tables 3 to 8 are
only available in electronic form at the CDS via anonymous ftp to
cdsarc.u-strasbg.fr (130.79.128.5) or via
http://cdsweb.u-strasbg.fr/Abstract.html. Figures A1 to A12 will be
made available only in the electronic version of the journal.}  of the
Hamburg/SAO Survey for Emission-Line Galaxies (HSS hereafter, SAO --
Special Astrophysical Observatory, Russia).  The list is a result of
the follow-up spectroscopy conducted with the 6\,m SAO RAS telescope
in 1998, 1999 and 2000.  The data of this snap-shot spectroscopy
survey confirmed 127 emission-line objects out of 176 observed
candidates and allowed their quantitative spectral classification.  We
could classify 76 emission-line objects as BCG/H{\sc ii} galaxies or
probable BCGs, 8 -- as QSOs, 2 -- as Seyfert galaxies, 2 -- as
super-associations in a subluminous spiral and an irregular galaxy,
and 37 as low-excitation objects -- either starburst nuclei (SBN), or
dwarf amorphous nuclei starburst galaxies (DANS). We could not
classify 2 ELGs. Furthermore, for 5 galaxies we did not detect any
significant emission lines.
For 91
emission-line galaxies, the redshifts and/or line intensities are
determined for the first time.  Of the remaining 28
previously known ELGs we give either improved data on the line
intensities or some independent measurements.

The candidates were taken from three different samples selected by 
different criteria. Among our first priority candidates we achieved 
a detection rate of emission-line objects (ELGs + QSOs)  of 68\%, 
among which 51\% are BCGs. Observations of a random selected sample
among our second priority candidates showed that only $\approx$10\% are
BCGs. We found that the confirmed BCGs have usually a blue colour
(($B-R$) $<$ 1\fm0) and a non-stellar appearance in the APM database.
Our third sample is comprised of second priority candidates fulfilling
these criteria derived from the APM. Follow-up spectroscopy of a
small subsample indicates that the expected detection rate for BCGs
is $\approx$40\%.


\end{abstract}

\section{Introduction}
The main goal of the Hamburg/SAO Survey for Emission-Line Galaxies
(HSS) is the search for emission-line galaxies (ELG) in order to
create a new deep sample of blue compact/H{\sc ii} galaxies (BCG) in a
large area of the sky with a size of the order 1700 square degrees.
Another important goal of this work is to search for new extremely
low-metallicity galaxies. The search is carried out on the objective
prism plates of the Hamburg Quasar Survey (Hagen et al.
\cite{Hagen95}).  The boundaries of our survey are $7^h 20^m$ to $17^h
40^m$ in right ascension and $+35\degr$ to $+50\degr$ in declination,
which bridges the gap between the zones of the Second Byurakan Survey
(SBS; Markarian et al.~\cite{Markarian83},
Stepanian~\cite{Stepanian94}) and the region covered by the Case
(Pesch et al. \cite{Pesch95}) survey.  The SBS is situated at $\alpha
= 7^h 40^m \div 17^h 20^m$, $\delta = +49\degr \div +61\degr$, while
the Case zone
corresponds to the region with $\alpha = 8^h 00^m \div 16^h 20^m$,
$\delta = +29\degr \div +38\degr$.  After combination of the four BCG
samples coming from the SBS (Izotov et al. \cite{Izotov93a},
\cite{Izotov93b}, Thuan et al. \cite{Thuan94}, Pustilnik et
al. \cite{Pustilnik95}), from the Case survey (Salzer et
al. \cite{Salzer95}, Ugryumov \cite{Ugryumov97}, Ugryumov et
al. \cite{Ugryumov98}), from the HSS, and from part of the Heidelberg
Void survey sample (Popescu et al. \cite{Popescu96}), a large Northern
BCG Sample
covering about 1 steradian will be available.

In the Papers I and II ((Ugryumov et al. \cite{Ugryumov99}, 
Pustilnik et al. \cite{Pustilnik99}) we presented results of
our survey in the region between $\delta=$40--50\degr, while
in Paper III (Hopp et al. \cite{PaperIII}) and the present paper
results from the strip $\delta=$35--40\degr\ are given. Forthcoming
papers will complete the follow-up spectroscopy in these two regions.

The article is organized as follows. In Section \ref{selection} we
will review our selection method and present the samples discussed
here. In Section \ref{Obs_red} we describe the spectroscopic
observations and the data reduction.  
In Section \ref{Res_follow} the results of the observations are
presented in several tables.  In Section \ref{Discussion} we briefly
discuss the new data and summarize the current state of the
Hamburg/SAO survey. Throughout this paper a Hubble constant H$_0$ = 75
km$\,$s$^{-1}$ Mpc$^{-1}$ is used.

\section{Selection method}
\label{selection}

The basic ideas of the HSS and its selection methods of ELG candidates
are described in Paper~I. The final selection was slightly modified to
improve significantly the detection rate of ELGs in the follow-up
spectroscopy as described in Paper~II. As it was outlined in Paper I,
the selection procedure provided us finally with two candidate lists
(first and second priorities): 1st -- objects showing a clear density
peak near $\lambda$~5000~\AA\ and blue continuum in the Hamburg Quasar Survey
objective-prism spectra scanned with high resolution; 2nd --
candidates with a blue continuum but without prominent emission
features or candidates with indications of emission peaks but with an
unusual continuum shape.  In short, the ELG candidate selection
criteria applied are a blue or flat continuum (near
$\lambda$~4000~\AA) and the presence of strong or moderate [O{\sc
iii}]\,$\lambda\lambda$\,4959,5007~\AA\ emission lines recognized on
digitized prism spectra. Candidates accepted had B-magnitudes in the
range $16^m - 19\fm5$.

Based on the experience with a training sample of BCGs drawn from the
Second Byurakan Survey (see Paper~I for details) the first priority
candidates were considered as highly probable H{\sc ii}/BCG type
emission galaxies. The follow-up snap-shot spectroscopy confirmed
that among all detected ELGs this type of galaxies constitutes up to
70--80\%. Thus, our main goal to create a large new sample of
BCG/H{\sc ii}-galaxies in the HSS region is achieved by follow-up
spectroscopy of the full sample of first priority candidates.

In this paper we will present the results for follow-up spectroscopy of
three samples listed in Table \ref{summary}.  The first sample is made
up by 139 first priority candidates in the strip $\delta=$35--40\degr\
of which 26 are known ELGs.  For the latter galaxies additional spectra
were required to improve their classification.

The other two samples were drawn from the second priority candidates
of the same strip. We found in Paper~I that the detection efficiency
of H{\sc ii}/BCG galaxies is rather low among them, prohibiting
follow-up spectroscopy of all second priority candidates. Moreover,
the second priority objects are about twice as numerous than the first
priority ones. At the faint end, this sample is also dominated by
candidates selected because of noise peaks in their objective prism
spectra. We created therefore a random selected sample of 43 second
priority objects from this strip matching in its magnitude
distribution the sample of first priority candidates. These objects
make up 10\% of the second priority sample in the magnitude range of
the first priority candidates.  We obtained follow-up spectroscopy of
all objects to study the general content and to determine the fraction
of BCGs in this sample.

The third sample (referred to as {\it APM Selected} in Table \ref{summary})
was created by applying additional selection criteria 
to the second priority candidates to increase the detection efficiency 
for BCG/H{\sc ii}-galaxies among them. These criteria and the results 
of follow-up spectroscopy are described in Section \ref{APMsel}.

\begin{table*}
\begin{center}
\caption[]{\label{summary} 
Summary of the samples observed and breakdown of the classifications after
follow-up spectroscopy
}
\begin{tabular}{llrrrrcrr}
\hline\noalign{\smallskip}
Candidate      & Sample      & N   & BCGs & Other & QSO &  Galaxies   & Stars & Not       \\
Classification &             &     &      & ELGs  &     &  without ELs&       & Classified\\
\hline\noalign{\smallskip}
First priority
 & new                       & 113 & 47   & 24    &  7  &  3       & 7     & 25        \\
 & already known             &  26 & 19   & 7     & --  & --       & --    & --        \\
 & Total                     & 139 & 66   & 31    &  7  &  3       & 7     & 25        \\[0.10cm]
\hline
Second priority
 & 6\,m observations   &  26 & 1    & 10    &  1  &  1       &  9    &  4        \\
 & 2.2\,m observations &  17 & 3    &  2    & --  &  1       &  3    &  8        \\
 & Random (total)            &  43 & 4    & 12    &  1  &  2       & 12    & 12        \\[0.10cm]
\hline
 & APM selected (total)      &  11 & 7    &  4    & --  & --       & --    & --        \\
\noalign{\smallskip}\hline\noalign{\smallskip}
\MC{2}{l}{Objects presented in this paper}
			     & 176 & 74   & 45    & 8   &  4       & 16    & 29      \\
\noalign{\smallskip}\hline
\end{tabular}
\end{center}
%
%
\begin{center}
\caption{\label{Tab1} Log of observations at the SAO 6\,m telescope}
\begin{tabular}{ccrllccc} \\ \hline
\MC{1}{c}{ Run } &
\MC{3}{c}{ Date } &
\MC{1}{c}{ Instrument } &
\MC{1}{c}{ Grating } &
\MC{1}{c}{ Wavelength } &
\MC{1}{c}{ Dispersion } \\

\MC{1}{c}{ No } & & & &  &
\MC{1}{c}{ [grooves/mm] } &
\MC{1}{c}{ Range [\AA] } &
\MC{1}{c}{ [\AA/pixel] } \\

\MC{1}{c}{ (1) } &
\MC{3}{c}{ (2) } &
\MC{1}{c}{ (3) } &
\MC{1}{c}{ (4) } &
\MC{1}{c}{ (5) } &
\MC{1}{c}{ (6) } \\
\hline
\\[-0.3cm]
\qq& 17--19 & Dec & 1998  & CCD, SP--124 & 300 & 3600--7800 & 4.6  \\
\qq& 08--13 & Feb & 1999  & CCD, LSS     & 325 & 3600--7800 & 4.6  \\
\qq&     02 & Sep & 1999  & CCD, LSS     & 650 & 3700--6100 & 2.4  \\
   &     04 & Sep & 1999  & CCD, LSS     & 325 & 3600--7800 & 4.6  \\
\qq&     02 & Feb & 2000  & CCD, LSS     & 650 & 3700--6100 & 2.4  \\
\hline \\[--0.2cm]
\end{tabular}
\end{center}
\end{table*}

\section{Spectral observations and data reduction}
\label{Obs_red}

\subsection{Observations}

All results presented below have been obtained by observations with
the Russian 6\,m telescope, mainly in the snap-shot mode during 4 runs
between December 1998 and February 2000. The spectrograph SP-124 attached to
the Nasmyth-1 focus of the telescope was used during the first run.
We used a grating with 300 grooves/mm (see log of observations
in Table~\ref{Tab1}) and a long slit of 2\arcsec$\times$40\arcsec.
The scale along the slit was 0.4\arcsec/pixel.

The Long-Slit Spectrograph (LSS in Table~\ref{Tab1}) (Afanasiev et al.
\cite{Afanasiev95}) attached to the telescope prime focus was used during
the remaining 3 runs. Most of the long-slit spectra
(1\farcs2--2\farcs0$\times$180\arcsec) were obtained with a grating of
325 grooves/mm, giving a dispersion of 4.6~\AA/pixel.  Additional data
were obtained with a grating of 650 grooves/mm giving a dispersion of
2.4~\AA/pixel.  The scale along the slit was 0.39\arcsec/pixel.  For
all observations we used the Photometrics CCD-detector PM1024 with
$24\times24\mu$m pixel size.

Normally, short exposures were used ($2-5$ minutes) in order to detect
strong emission lines, to measure redshifts and make a first
classification.  Reference spectra of an Ar--Ne--He lamp were recorded
before or after each observation to provide a wavelength calibration.
Spectrophotometric standard stars from Oke (\cite{Oke90}) and Bohlin
(\cite{Bohlin96}) were observed for flux calibration at least twice a
night.  All observations and data acquisition have been conducted
under the {\tt NICE} software package by Kniazev \& Shergin (1995) in
the MIDAS\footnote{MIDAS is an acronym for the European Southern
Observatory package --- Munich Image Data Analysis System.}
environment.

Part of the second priority candidates was observed with the 2.2m telescope
on Calar Alto in June 1999. These observations will be presented in a 
forthcoming paper, although we will make use of the results in our analysis
below.
 
\subsection{Data reduction}

The data reduction was done at SAO with the IRAF\footnote{IRAF is
distributed by National Optical Astronomical Observatories, which is
operated by the Association of Universities for Research in Astronomy,
Inc., under cooperative agreement with the National Science
Foundation} and the MIDAS software packages.  In all details of the
reduction process and the measuring of line parameters we followed the
procedures described in Paper\,III.  Since we
present a substantial number of objects with redshifts known from
earlier publications, we could independently test the quality of our
wavelength calibration.  The results of these tests indicate that our
internal errors $\sigma_V$ shown in Table~\ref{Tab2} are close to the external
errors and do not change from run to run.

\section{Results of follow--up spectroscopy}
\label{Res_follow}

\subsection{First priority candidates}

In total 139 first priority candidates have been observed (Table
\ref{summary}).  Of them 26 objects were known as ELG before. In
particular, seven of them are from our Paper III, for which the
classification was either unknown or uncertain.
Of these 139 objects
97 (75\%) are new or confirmed emission-line galaxies.

\subsubsection{Emission-line galaxies}

The emission line galaxies observed are listed in Table~\ref{Tab2} containing
the following information: \\
 {\it column 1:} The object's IAU-type name with the prefix HS.\\ 
 {\it column 2:} Right ascension for equinox B1950. \\
 {\it column 3:} Declination for equinox B1950.
The coordinates were measured on direct plates of the HQS
and are accurate to $\sim$ 2$\arcsec$ (Hagen et al. \cite{Hagen95}). \\
 {\it column 4:} Heliocentric velocity and its r.m.s. uncertainty in
km~s$^{-1}$. \\
 {\it column 5:} Apparent B-magnitude obtained by calibration of the digitized
photoplates with photometric standard stars (Engels et al. \cite{Engels94}),
having an r.m.s. accuracy of $\sim$ $0\fm5$ for objects fainter than
m$_{\rm B}$ = $16\fm0$ (Popescu et al. \cite{Popescu96}).
Since the algorithm to calibrate the objective prism spectra is
optimized for point sources the brightnesses of extended galaxies are
underestimated. The resulting systematic uncertainties are expected to
be as large as 2 mag (Popescu et al. \cite{Popescu96}). For about 1/3
of our objects, B-magnitudes are unavailable at the moment. We present
for them blue magnitudes obtained from the APM database. They are
marked by a ``plus" before the value in the corresponding
column. According to our estimate they are systematically brighter by
$0\fm92$ than the B-magnitudes obtained by calibration of the
digitized photoplates (r.m.s.  $1\fm02$).
For all objects marked as from Popescu et al. (\cite{Popescu99}) one may
find improved B-magnitudes in Vennik, Hopp \& Popescu (\cite{Vennik2000})
which we do not list here for the sake of homogeneity.\\
 {\it column 6:} Absolute B-magnitude, calculated from the apparent
B-magnitude and the heliocentric velocity. No correction for galactic
extinction is made as all objects are located at high
galactic latitudes and because the corrections are significantly smaller
than the uncertainties of the magnitudes. \\
 {\it column 7:} Preliminary spectral classification type according to
the spectral data presented in this article. BCG means that the galaxy
possesses a characteristic H{\sc ii}-region spectrum and that the
luminosity is low enough. SBN and DANS are galaxies of lower
excitation with a corresponding position in line ratio diagrams, as
discussed in Paper~I. SBN are the brighter fraction of this type. We
here follow the notation of Salzer et al. (\cite{Salzeretal89}).
Seyfert galaxies are separated mainly on diagnostic diagrams as AGN.
The criterion of broad lines was also used for the Sy classification.
With SA two probable super-association in a spiral 
and an irregular galaxy are denoted. Two ELGs are difficult to classify.
They are coded as NON. \\
 {\it column 8:} One or more alternative names, according to the
information from NED.\footnote{http://nedwww.ipac.caltech.edu/}
References to other sources of spectral information indicating that a 
galaxy is an ELG are given in bold face.

The spectra of all emission-line galaxies are shown in Appendix~A,
which is available only in the electronic version of the journal.

The results of line flux measurements are given in Table~\ref{Tab3}.
It contains the following information: \\
 {\it column 1:} The object's IAU-type name with the prefix HS.
By asterisk we note the objects observed  during
non-photometric conditions. \\
 {\it column 2:} Observed flux (in
10$^{-16}$\,erg\,s$^{-1}$\,cm$^{-2}$) of the H$\beta$\, line.
For few objects without H$\beta$ emission line the fluxes are given
for H$\alpha$ and marked by a ``plus''.
For several objects observed during non-photometric
conditions this parameter is unreliable and marked by (:).   \\
 {\it columns 3,4,5:} The observed flux ratios [O{\sc ii}]/H$\beta$,
[O{\sc iii}]/H$\beta$ and H$\alpha$/H$\beta$.
For few objects without H$\beta$ flux  ratios are given
for H$\alpha$ and marked by a ``plus''.\\
 {\it columns 6,7:} The observed flux ratios
[N{\sc ii}]\,$\lambda$\,6583~\AA/H$\alpha$, and
([S{\sc ii}]\,$\lambda$\,6716~\AA\ + \,$\lambda$\,6731~\AA)/H$\alpha$. \\
 {\it columns 8,9,10:} Equivalent widths of the lines
[O{\sc ii}]\,$\lambda$\,3727~\AA, H$\beta$ and
[O{\sc iii}]\,$\lambda$\,5007~\AA.
For few objects without detected H$\beta$ emission line the equivalent
widths are given for H$\alpha$ and marked by a ``plus''.

Among the 97 ELGs observed as first-priority candidates,
66 are classified as BCGs or
probable BCGs.  Two very faint objects (HS~1134+3640 and HS~1308+3845)
are probably super-associations in the dwarf spiral NGC~3755 and
in the Im galaxy UGC~8261. Two ELGs are probable LINERs.
One candidate is difficult to classify. The
remaining 26 ELGs are objects of lower excitation: either starburst
nuclei galaxies (SBN and probable SBN) or their lower mass analogs
dwarf amorphous nuclear starburst galaxies (DANS or probable DANS).

\noindent
Below we give notes on several individual objects: 

\noindent
{\it HS~0847+3639}: The HS magnitude for this galaxy seems to be too
faint. The KUG magnitude (Takase \& Miyauchi-Isobe \cite{Takase1993})
B=15\fm4  corresponds to M$_B$=--19\fm6 and an SBN
classification as given in Table~\ref{Tab2}. \\
{\it HS~0934+3629}: The FWHMs, corrected for instrumental resolution,
for H$\alpha$ and H$\beta$ are $\approx$1800 km~s$^{-1}$ and
$\approx$1400  km~s$^{-1}$, respectively. \\
{\it HS~1047+3714}: Very strong N{\sc ii} line $\lambda$6583~\AA, H$\beta$
  is only seen in absorption.\\
{\it HS~1116+3951}: Uncertain H$\alpha$/H$\beta$ ratio because of a cosmic
hit on H$\alpha$.\\
{\it HS~1134+3639}: Probable low-mass companion of the  galaxy NGC~3755 (the 
distance is $\approx$230\arcsec\ or $\sim$25 kpc). \\
{\it HS~1134+3640}: Possible giant H{\sc ii}-region at the very edge of
the  SAB(rs)c pec galaxy NGC~3755 with V$_{hel}=$1570 km~s$^{-1}$, seen
  at an inclination angle of $\approx$60$^{\circ}$.
The HI-line width at the level of 0.2 of the peak flux value W$_{0.2}$=290
  km~s$^{-1}$ (Huchtmeier \& Richter~\cite{Hucht89}) corresponds
  to a maximum V$_{rot}$ of $\approx$150 km~s$^{-1}$.
  In accordance with the difference in  radial velocities (--108 km~s$^{-1}$)
of HS~1134+3640 and the dynamical center of NGC~3755, 
HS~1134+3640 is either
an H{\sc ii}-region in NGC~3755, or a companion like HS~1134+3639. 
The real
situation can be checked only if it is determined whether this edge of 
NGC~3755,
corresponds to the receding or approaching spiral arm. \\
{\it HS~1308+3545}: Giant H{\sc ii}-region at the edge of the B=16\fm0 Im 
galaxy
   UGC~8261 with V$_{hel}=$852 km~s$^{-1}$. Its HI-line width 
W$_{0.2}$=127 km~s$^{-1}$ corresponds to
   a maximum V$_{rot}$ of $\approx$65 km~s$^{-1}$. This is consistent
   with the measured difference in the radial velocities of HS~1308+3545 and
the   dynamical center of the Im galaxy. \\
{\it HS~1620+4003}: The profiles of the O{\sc iii}-lines $\lambda$ 4959,5007~\AA\
have a composite structure with a narrow (FWHM$_{5007}$ = FWHM$_{H\beta}$) and
broad (FWHM = $\approx$1800 km~s$^{-1}$) component. The broad to narrow
component flux ratio is 1.44.

\subsubsection{Quasars}

In the course of our follow-up spectroscopy, seven QSOs were
discovered with a strong emission line in the wavelength region
between 5000~\AA\ and the sensitivity break of the Kodak IIIa-J
photoemulsion near 5400~\AA.  In all of them, we identified
Ly$\alpha$\,$\lambda$\,1216 redshifted to z $\sim$ 3 as the
responsible line.
These mostly faint quasars are not found by the HQS
itself, which focuses on objects with brightnesses B$\le$17\fm5
(Hagen et al. \cite{Hagen99}).
The data for these 7 new high-redshift quasars (and
one from the second priority candidates) are presented in
Table~\ref{Tab4}.  Finding charts and plots of their spectra can be
found on www-site of the Hamburg Quasar Survey
(http://www.hs.uni-hamburg.de/hqs.html).

\subsubsection{Absorption-line galaxies}

For three bright non-ELG galaxies (and one from the second priority
candidates) the signal-to-noise ratio of our spectra was sufficient to
detect absorption lines, allowing the determination of redshifts. The
data are presented in Table~\ref{Tab5}.

\subsubsection{Stellar objects}

To separate the stars among the objects missing detectable emission
lines we cross-correlated a list of the most common stellar features
with the observed spectra.  In total, 7 objects with definite
stellar spectra and redshifts close to zero were identified (and 9
from the second priority candidates).  They were classified roughly in
categories from definite A-stars to G-stars, with most of them
intermediate between A and F. The data for these stars are presented
in Table~\ref{Tab6}.

\subsubsection{Not-classified  objects}

Twenty five non emission-line objects are hard to classify at
all. Their continua have too low signal-to-noise ratio to detect
trustworthy absorption features, or the EWs of the emission lines are
too small.

\begin{figure}
   \hspace*{-1.0cm}
   \psfig{figure=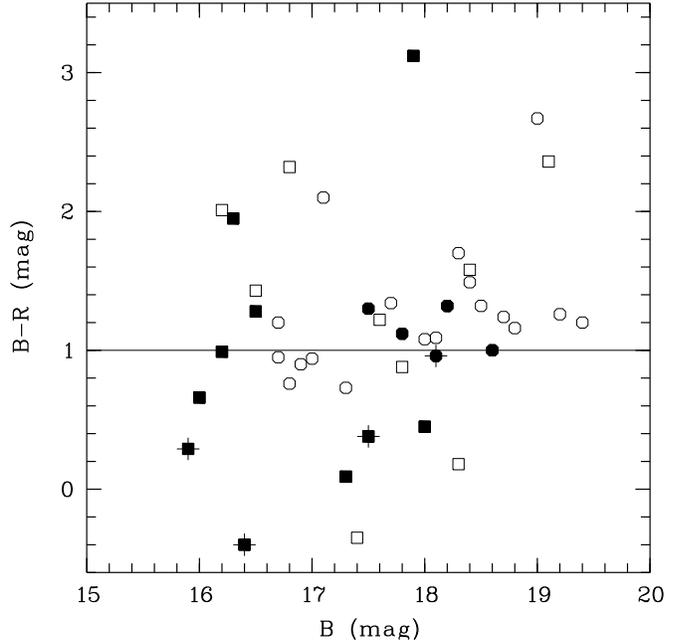,angle=-90,width=13cm}
   \caption{%
   Colour distribution for the random sample of 42 second priority
   candidates in the strip $\delta$=35--40\degr\
   (HS~1423+3945 is not included because
   its APM-colour is not available).
   The APM classification is shown by circles for stellar
   and squares for non-stellar images. The filled symbols show the
   emission-line objects identified by follow-up spectroscopy.
   BCG/H{\sc ii}-galaxies are marked by an additional cross. 
   4 of 5 APM stellar-type
   emission-line objects are either QSOs or distant luminous galaxies.
   The horizontal line at ($B-R$) = 1\fm0 shows the colour limit for the
   selection of BCG candidates among
   the second priority objects classified as non-stellar in the APM.
}
     \label{fig:HS2_II_sub_colours}
\end{figure}

\subsection {Second priority candidates}
\label{Second_prior}

\subsubsection{The random selected sample}
\label{test}

From the random selected subsample of second priority candidates we
observed twenty six objects during the 6-m telescope runs (mainly
those with R.A.$<$ 11$^h$) along with the observations of first
priority targets.  Altogether we found 12 emission-line objects: one
appeared to be a QSO with $z =$3.15, and the rest are various types of
ELGs. They are shown in
Table~\ref{Tab2} and ~\ref{Tab3}, respectively, and are marked by
$\dagger$.  Nine objects are stars of spectral classes from A to G,
with one M-star. One object is an absorption-line galaxy.  The
remaining four second priority candidates have no (trustworthy)
emission lines, and are probably either absorption line galaxies, or
various types of stars. They are considered as ``not classified'' in
Table \ref{summary}.  Among the 11 ELGs only one has a spectrum and an
absolute magnitude indicative of a BCG/H{\sc ii}-galaxy. The others
are either low excitation ELGs like SBN and DANS, or AGN like LINERs
or Sy galaxies.

The remaining 17 candidates were observed
with the 2.2m telescope and three probable BCGs were found among them.
Preliminary spectral classification type for these objects are shown
in the last column of Table~\ref{Tab3a}. Classification of ELGs follows
that described for Table~\ref{Tab2}. ``Star'' is star-like object with absorptions
at zero redshift. ``ABS'' is a galaxy with detected and identified absorption lines.
``NOEM'' is an object with no (trustworthy) emission lines detected.

In total, we found 4 (probable) BCG/H{\sc ii}-galaxies
in the random selected sample making up a fraction of $\approx$10\%.

\subsubsection{The APM selected sample}
\label{APMsel}

To improve the efficiency for the selection of BCGs among the second
priority candidates we introduced additional selection criteria using
the APM (Irwin \cite{Irwin98}) database. 
An analysis of the spectroscopic classifications of
all objects from the random subsample
resulted in the following conclusions:
\begin{itemize}
\item The second priority candidates classified as BCG/H{\sc ii} based on
 their slit spectra all have rather blue colours according
to the APM database (APM $(B-R) \le 1\fm0$). In this colour range BCGs
constitute about 33\% of all second priority objects classified by APM as 
galaxies (Figure \ref{fig:HS2_II_sub_colours}),
\item Most of the ELGs of other types are redder than BCGs,
\item One BCG
of the four newly found is classified in the APM
as stellar.  Possibly at faint magnitudes some fraction of very
compact BCGs are not distinguished by the APM from blue stars.
Therefore also the APM information is of no help
to discriminate the very compact faint BCGs from blue stars
at the faint end of the second priority candidates.
\end{itemize}

Therefore, having in mind to develop a strategy to pick up as many as
possible low-mass BCG/H{\sc ii}-galaxies in the HSS region, but
accounting as well for a high enough detection rate, we created an
subsample of second priority BCG candidates, which fulfill the
following additional criteria:
\begin{itemize}
\item APM non-stellar classification, and 
\item a blue colour ($(B-R)< 1\fm0$) on the APM
\end{itemize}

Of 633 objects from the second priority list in the strip
$\delta$=35--40\degr\  80 candidates were selected by these criteria.
Eleven candidates from this additional list were observed by 
snap-shot spectroscopy. All of them have turned out to be
ELGs of various types. 3 of them are certain BCGs, and 4 more are
probable BCG/H{\sc ii}-galaxies. They are listed in Table~\ref{Tab2}
and are  marked with $\ddagger$.

\section{Discussion}
\label{Discussion}

\subsection{The Fourth List}

As shown in Table~\ref{summary}, among 104 observed emission-line
objects 66 (63\%) were classified based on the character of
their spectra and their absolute magnitudes as H{\sc ii}/BCGs or
probable BCGs.  Since the main goal of the HSS is an efficient search
for new BCGs, the fraction of this type among all confirmed
ELGs of the first priority list
($\sim$68~\%, or 63~\% among all emission-line objects)
is encouraging.

The distribution of  new HSS ELGs in the line-ratio diagrams
[O{\sc iii}]\,$\lambda$\,5007/H$\beta$ versus
[N{\sc ii}]\,$\lambda$\,6583/H$\alpha$ and
[O{\sc iii}]\,$\lambda$\,5007/H$\beta$ versus
[O{\sc ii}]\,$\lambda$\,3727/[O{\sc iii}]\,$\lambda$\,5007
(see Baldwin et al. (\cite{Baldwin81}), Veilleux \& Osterbrock
(\cite{Veilleux87}) for details) in general is similar
to that shown in Paper~I
as may be expected since the selection criteria are identical.
Several new BCGs with very strong emission lines located in the
metal-poor regions of these diagrams
were reobserved later with
higher signal-to-noise ratio, and four
of them (0951+3841, 1028+3843, 1124+3635
and 1309+3806) are found to have O/H $< $ 1/10 (O/H)$_{\odot}$.
A full description of the selection procedure for low-metallicity candidates
based on snap-shot spectra is given in Kniazev et al. (\cite{Kniazev2000}).

The snap-shot spectroscopy of the random selected sample of
second priority candidates detected only 4 additional BCG/HII
galaxies ($\approx$10\%, cf. Table~\ref{summary}).
We expect therefore, not much more than $\approx$60 BCGs
among the 633 second priority candidates
of the $\delta$=35--40\degr\ strip.

To find these BCGs efficiently we used additional selection criteria
based on the APM, which are fulfilled also by most of the BCGs from
the first priority list.
For this subsample of 80 candidates
spectroscopic information is available for 24 objects: 11 were observed
by us due to this selection, 10 were already observed as part of the random
sample, and for 3 the information was taken from the literature. Altogether
10 BCGs or probable BCGs were found among them.
The current estimate of detection
rate for BCGs in the APM selected sample is therefore $\approx$40\%.

\subsection{Summary of the present status of the survey}

Altogether among the objects of first priority in Papers I through IV,
we discovered 321 new
emission-line objects (20 of them QSOs), and for 55 known
ELGs we got quantitative data for their emission lines.
Preliminary classification of the 356
ELGs yields 275 ($\sim$77~\%) confident or probable
blue compact/low-mass H{\sc ii}-galaxies.
This large fraction demonstrates the high
efficiency of this survey to find low-mass galaxies with H{\sc ii}-type
spectra on the Hamburg Quasar Survey photoplates. 
A statistical analysis of
this BCG sample, supplemented with galaxies from the SBS, the Case, and the
Heidelberg Void samples, is underway.
Fourteen more BCGs were found among the second priority candidates.

\section{Conclusions}

We conducted follow-up spectroscopy of candidates from the Hamburg/SAO
Survey for ELGs.
Summarizing the results presented, the analysis of the content of various
types of objects, and the discussion above, we draw the following
conclusions:

\begin{itemize}

\item The applied methods to detect ELG candidates on the plates of the
      Hamburg Quasar Survey give a reasonably high detection rate of
      emission-line objects:
      $\sim$~77\% in the average, among the first priority candidates
      as defined in Papers I to IV.

\item The high fraction of BCG/H{\sc ii} galaxies among all newly
	discovered ELGs (about 68~\% in this paper)
	is in line with our main
	goal --- to pick up efficiently a deep BCG sample in the sky region 
        under analysis.

\item Besides of ELGs we found also 8 new quasars, all with Ly$\alpha$
      in the wavelength region $4950-5100$~\AA\, (i.e with 3.15 $< z < $ 3.30)
      near the red boundary of the IIIa-J objective prism photoplates.
      These objects are a byproduct of the survey, as their Ly$\alpha$ line 
      is mistaken for the [O{\sc iii}]\,$\lambda$\,5007~\AA\ line.

\item The snap-shot spectroscopy of the random selected sample of
      second priority candidates shows that BCG/H{\sc ii}-galaxies
      represent only a small part of the second priority sample
      ($\approx$10\%).  Applying additional
      selection criteria based on APM classification and $(B-R)$ colour
      allows to extract more reliably these BCGs, except for 
      very compact ones.
\end{itemize}

\begin{acknowledgements}

This work was supported by the grant of the Deutsche
Forschungsgemeinschaft No.~436 RUS~17/77/94. U.A.V. is very grateful
to the staff of the Hamburg Observatory for their hospitality and kind
assistance.  Support by the INTAS grant No.~96-0500 was crucial to
proceed with the Hamburg/SAO survey declination band centered on
+37.5$^{\circ}$. We notice that the use of APM facility was extremely valuable
for selection methodics of additional candidates to BCGs from the 2-nd
priority list.  The authors are grateful to the referee G.Comte for his
useful advises, allowing to improve the presentation of results.
This research has made use of the NASA/IPAC Extragalactic
Database (NED) which is operated by the Jet Propulsion Laboratory,
California Institute of
Technology, under contract with the National Aeronautics and Space Administration.

\end{acknowledgements}

\clearpage



\scriptsize
\setcounter{qub}{0}

\begin{table*}[h]

\begin{center}
\caption{\label{Tab2} Coordinates, Velocities and Magnitudes of
Emission--Line Galaxies}

\begin{tabular}{rlllrrcll} \hline \\[-0.35cm]
\multicolumn{1}{c}{\#}               &
\multicolumn{1}{c}{Name}             &
\multicolumn{1}{c}{$\alpha\,(1950)$} &
\multicolumn{1}{c}{$\delta\,(1950)$} &
\multicolumn{1}{c}{$V_{hel}^{\,a}\pm\sigma_V$}    &
\multicolumn{1}{r}{m$_{B}$}          &
\multicolumn{1}{c}{M$_{B}^{\;\;b}$}  &
\multicolumn{1}{c}{Type}             &
\multicolumn{1}{l}{Other names from NED}\\
&&&&&&&& \multicolumn{1}{c}{and number of reference}\\
&
\multicolumn{1}{c}{ (1) } &
\multicolumn{1}{c}{ (2) } &
\multicolumn{1}{c}{ (3) } &
\multicolumn{1}{c}{ (4) } &
\multicolumn{1}{c}{ (5) } &
\multicolumn{1}{c}{ (6) } &
\multicolumn{1}{c}{ (7) } &
\multicolumn{1}{c}{ (8) } \\
\hline
\qq& HS 0732+3529            &07 32 23.3 &+35 29 07  &3730  $\pm$ 42    &17.80 &--15.68&BCG      & \\
\qq& HS 0758+3901            &07 58 58.2 &+39 01 29  &11621 $\pm$ 57    &19.20 &--16.75&BCG      & \\
\qq& HS 0759+3811            &07 59 17.0 &+38 11 51  &6142  $\pm$ 54    &17.80 &--16.77&BCG      & \\
\qq& HS 0801+3740            &08 01 51.5 &+37 40 21  &9823  $\pm$ 54    &18.70 &--16.89&BCG      & \\
\qq& HS 0801+3746$^\dagger$  &08 01 47.5 &+37 46 06  &10793 $\pm$ 100   &16.20 &--19.59&NON      & \\
\qq& HS 0803+3648            &08 03 29.2 &+36 48 38  &9698  $\pm$ 66    &18.40 &--17.16&BCG      & \\
\qq& HS 0826+3521            &08 26 52.2 &+35 21 17  &18910 $\pm$ 200   &19.20 &--17.81&NON      & \\
\qq& HS 0834+3741            &08 34 11.6 &+37 41 14  &12509 $\pm$ 15    &18.30 &--17.81&BCG?     & \\
\qq& HS 0837+3816            &08 37 53.4 &+38 16 08  &12269 $\pm$ 3     &17.80 &--18.27&SBN?     & \\
\qq& HS 0840+3938            &08 40 18.5 &+39 38 21  &9646  $\pm$ 102   &17.40 &--18.15&BCG?     & \\
\qq& HS 0840+4013            &08 40 42.4 &+40 13 23  &8887  $\pm$ 15    &17.40 &--17.97&DANS?    & KUG 0840+402 \\
\qq& HS 0841+3959            &08 41 31.0 &+39 59 15  &34316 $\pm$ 90    &+18.10&--20.20&SBN      & \\
\qq& HS 0843+3637            &08 43 23.6 &+36 37 23  &3207  $\pm$ 51    &15.40 &--17.76&BCG      &Mkn 627,CG 224, {\bf 1} \\
\qq& HS 0847+3639            &08 47 29.5 &+36 39 17  &7659  $\pm$ 108   &18.10 &--16.95&SBN?     & KUG 0847+366, {\bf 7} \\
\qq& HS 0852+3614            &08 52 00.8 &+36 14 54  &8123  $\pm$ 72    &19.20 &--15.97&BCG?     & \\
\qq& HS 0852+4003            &08 52 53.2 &+40 03 41  &1941  $\pm$ 20    &16.50 &--15.56&BCG      & KUG 0852+400A \\
\qq& HS 0900+3651            &09 00 55.0 &+36 51 10  &17303 $\pm$ 48    &18.40 &--18.42&BCG      & \\
\qq& HS 0902+3759            &09 02 37.5 &+37 59 39  &14331 $\pm$ 60    &19.30 &--17.11&SBN      &IRAS 09026+3759 \\
\qq& HS 0902+4005            &09 02 58.3 &+40 05 38  &14453 $\pm$ 27    &17.80 &--18.62&BCG      & \\
\qq& HS 0903+4349            &09 03 31.9 &+43 49 45  &7611  $\pm$ 100   &20.20 &--14.83&BCG      & \\
\qq& HS 0904+4357            &09 04 12.7 &+43 57 31  &14814 $\pm$ 74    &18.70 &--17.78&DANS?    & \\
\qq& HS 0904+4400            &09 04 10.1 &+44 00 56  &17316 $\pm$ 55    &17.40 &--19.42&BCG?     & \\
\qq& HS 0905+3948            &09 05 45.2 &+39 48 24  &12850 $\pm$ 21    &18.60 &--17.57&BCG      & \\
\qq& HS 0907+3728            &09 07 02.4 &+37 28 08  &9045  $\pm$ 30    &16.40 &--19.01&SBN      & \\
\qq& HS 0913+3516            &09 13 59.1 &+35 16 48  &17098 $\pm$ 63    &19.50 &--17.29&BCG      & \\
\qq& HS 0925+3529            &09 25 41.4 &+35 29 47  &4556  $\pm$ 15    &19.20 &--14.72&BCG      & \\
\qq& HS 0928+3725$^\dagger$  &09 28 24.7 &+37 25 25  &14788 $\pm$ 39    &18.00 &--18.47&SBN?     & \\
\qq& HS 0928+4006            &09 28 35.6 &+40 06 57  &4295  $\pm$ 33    &18.90 &--14.89&BCG      & \\
\qq& HS 0930+3548            &09 30 08.3 &+35 48 41  &6871  $\pm$ 30    &18.50 &--16.31&BCG      & \\
\qq& HS 0934+3629$^\dagger$  &09 34 00.3 &+36 29 07  &53840 $\pm$ 66    &17.50 &--21.78&Sy1?     &IRAS F09339+3629 \\
\qq& HS 0936+3601$^\dagger$  &09 36 03.9 &+36 01 40  &6011  $\pm$ 37    &17.30 &--17.22&DANS?    & \\
\qq& HS 0936+3648$^\dagger$  &09 36 19.7 &+36 48 04  &6018  $\pm$ 39    &16.00 &--18.52&SBN?     &CG270, {\bf 4} \\
\qq& HS 0938+3544$^\dagger$  &09 38 45.7 &+35 44 18  &56334 $\pm$ 61    &18.60 &--20.78&Sy1?     & \\
\qq& HS 0942+3600            &09 42 09.4 &+36 00 30  &6668  $\pm$ 51    &16.20 &--18.54&BCG      & CG 278, {\bf 10} \\
\qq& HS 0942+3644            &09 42 41.2 &+36 44 58  &9856  $\pm$ 36    &15.80 &--19.79&SBN      & CG 283 \\
\qq& HS 0947+3559            &09 47 33.7 &+35 59 55  &5960  $\pm$ 54    &16.30 &--18.20&SBN      &CG295, KUG 0947+359, {\bf 1} \\
\qq& HS 0951+3606            &09 51 24.7 &+36 06 29  &12023 $\pm$ 59    &17.50 &--18.52&SBN?     & \\
\qq& HS 0951+3841            &09 51 46.4 &+38 41 13  &5172  $\pm$ 10    &16.80 &--17.39&BCG      & \\
\qq& HS 0952+3803            &09 52 39.9 &+38 03 27  &9535  $\pm$ 106   &18.20 &--17.32&DANS?    & \\
\hline
\end{tabular}
\end{center}
\end{table*}

\newpage

\begin{table*}[h]

\begin{center}
\flushleft {\bf Table 3.} (Continued)

\begin{tabular}{rlllrrcll} \\[-0.22cm] \hline \\[-0.35cm]
\multicolumn{1}{c}{\#}               &
\multicolumn{1}{c}{Name}             &
\multicolumn{1}{c}{$\alpha\,(1950)$} &
\multicolumn{1}{c}{$\delta\,(1950)$} &
\multicolumn{1}{c}{$V_{hel}^{\,a}\pm\sigma_V$}    &
\multicolumn{1}{r}{m$_{B}$}          &
\multicolumn{1}{c}{M$_{B}^{\;\;b}$}  &
\multicolumn{1}{c}{Type}             &
\multicolumn{1}{l}{Other names from NED}\\
&&&&&&&& \multicolumn{1}{c}{and number of reference}\\
&
\multicolumn{1}{c}{ (1) } &
\multicolumn{1}{c}{ (2) } &
\multicolumn{1}{c}{ (3) } &
\multicolumn{1}{c}{ (4) } &
\multicolumn{1}{c}{ (5) } &
\multicolumn{1}{c}{ (6) } &
\multicolumn{1}{c}{ (7) } &
\multicolumn{1}{c}{ (8) } \\
\hline
\qq& HS 0958+4043            &09 58 23.4 &+40 43 45  &13493 $\pm$ 56    &19.50 &--16.78&BCG      & \\
\qq& HS 1014+4030            &10 14 26.2 &+40 30 00  &19986 $\pm$ 46    &17.40 &--19.73&BCG      & \\
\qq& HS 1014+4916            &10 14 01.9 &+49 16 47  &15904 $\pm$ 29    &18.00 &--18.63&SBN?     & \\
\qq& HS 1018+3847            &10 18 49.6 &+38 47 58  &4204  $\pm$ 24    &16.50 &--17.24&BCG      & \\
\qq& HS 1021+3637$^\dagger$  &10 21 57.3 &+36 37 06  &52551 $\pm$ 108   &18.20 &--21.03&SBN?     & \\
\qq& HS 1021+4218            &10 21 12.8 &+42 18 36  &11000 $\pm$ 20    &18.00 &--17.83&BCG      & \\
\qq& HS 1025+4005$^\dagger$  &10 25 51.8 &+40 05 04  &9040  $\pm$ 45    &15.90 &--19.51&BCG      & KUG 1025+400D \\
\qq& HS 1026+3521            &10 26 25.7 &+35 21 02  &11129 $\pm$ 30    &18.20 &--17.66&BCG      & \\
\qq& HS 1028+3843            &10 28 57.7 &+38 43 34  &9014  $\pm$ 21    &19.40 &--16.00&BCG      & \\
\qq& HS 1047+3714$^\dagger$  &10 47 13.6 &+37 14 23  &10461 $\pm$ 45    &16.30 &--19.42&LINER?   & \\
\qq& HS 1048+4026            &10 48 00.8 &+40 26 08  &38840 $\pm$ 51    &18.90 &--19.67&SBN?     & \\
\qq& HS 1050+4002            &10 50 40.4 &+40 02 35  &8451  $\pm$ 45    &17.90 &--17.36&BCG      & \\
\qq& HS 1059+3934            &10 59 22.0 &+39 34 54  &3274  $\pm$ 66    &18.20 &--15.00&BCG      & \\
\qq& HS 1102+3644            &11 02 05.1 &+36 44 50  &22012 $\pm$ 204   &17.00 &--20.34&LINER?   & \\
\qq& HS 1107+3524            &11 07 53.1 &+35 24 36  &9492  $\pm$ 18    &17.40 &--18.11&BCG      & {\bf 3}\\
\qq& HS 1107+3637            &11 07 04.0 &+36 37 16  &8088  $\pm$ 15    &19.50 &--15.66&BCG      & {\bf 3}\\
\qq& HS 1107+3831            &11 07 09.7 &+38 31 14  &9543  $\pm$ 33    &+14.34&--21.18&BCG      &CG1388, KUG 1107+385 \\
\qq& HS 1116+3951            &11 16 33.9 &+39 51 58  &10090 $\pm$ 131   &+16.93&--18.71&SBN      & \\
\qq& HS 1118+3842            &11 18 46.8 &+38 42 06  &10433 $\pm$ 75    &+18.52&--17.20&BCG      & \\
\qq& HS 1124+3635            &11 24 49.4 &+36 35 59  &9160  $\pm$ 21    &+17.96&--17.43&BCG      & \\
\qq& HS 1125+3624            &11 25 43.9 &+36 24 23  &10488 $\pm$ 39    &+18.31&--17.42&BCG      & \\
\qq& HS 1125+3748            &11 25 22.4 &+37 48 35  &10689 $\pm$ 18    &+16.50&--19.27&BCG      &CSO 343 \\
\qq& HS 1126+3701            &11 26 33.8 &+37 01 27  &12808 $\pm$ 50    &+16.02&--20.16&BCG      & \\
\qq& HS 1129+3730            &11 29 11.2 &+37 30 14  &12787 $\pm$ 33    &+17.71&--18.45&BCG      & \\
\qq& HS 1134+3639            &11 34 15.7 &+36 39 52  &1595  $\pm$ 36    &17.10 &--14.54&BCG      & NPM1G +36.0258 \\
\qq& HS 1134+3640            &11 34 00.5 &+36 40 19  &1462  $\pm$ 58    &      &       &SA in Sp?&SA in NGC 3755 \\
\qq& HS 1135+3709            &11 35 38.0 &+37 09 58  &3251  $\pm$ 49    &18.10 &--15.08&DANS     & \\
\qq& HS 1144+3552            &11 44 37.6 &+35 52 38  &11001 $\pm$ 33    &+17.16&--18.67&BCG      & \\
\qq& HS 1145+3709            &11 45 53.4 &+37 09 55  &12129 $\pm$ 39    &16.60 &--19.44&SBN      &CSO 365, KUG 1145+371, {\bf 1} \\
\qq& HS 1147+3653            &11 47 53.3 &+36 53 14  &9207  $\pm$ 43    &+17.80&--17.65&DANS?    & \\
\qq& HS 1150+3756            &11 50 14.8 &+37 56 04  &11183 $\pm$ 90    &+19.08&--16.79&BCG?     &PC 1150+3756, {\bf 8} \\
\qq& HS 1159+3701            &11 59 34.6 &+37 01 29  &6847  $\pm$ 67    &17.40 &--17.40&BCG      & \\
\qq& HS 1221+3721            &12 21 21.1 &+37 21 56  &3907  $\pm$ 27    &17.90 &--15.68&BCG      & \\
\qq& HS 1228+3632            &12 28 04.7 &+36 32 12  &10329 $\pm$ 54    &+19.30&--16.40&BCG      & \\
\qq& HS 1240+3902            &12 40 52.9 &+39 02 34  &7030  $\pm$ 30    &+17.02&--17.84&BCG      & \\
\qq& HS 1300+3646            &13 00 24.1 &+36 46 31  &9081  $\pm$ 48    &+18.41&--17.01&BCG      & NGP9 F269-0522706 \\
\qq& HS 1301+3835            &13 01 26.8 &+38 35 02  &8732  $\pm$ 45    &17.40 &--17.93&BCG?     & \\
\qq& HS 1308+3545            &13 08 43.3 &+35 45 55  &776   $\pm$ 54    &18.10 &--11.97&SA in Im & UGC 8261 \\
\qq& HS 1309+3806            &13 09 17.9 &+38 06 54  &16011 $\pm$ 36    &+19.23&--17.42&BCG      & NGP9 F269-0162165 \\
\qq& HS 1318+3624            &13 18 34.4 &+36 24 48  &7590  $\pm$ 45    &18.00 &--17.03&BCG      & \\
\hline
\end{tabular}
\end{center}
\end{table*}

\newpage

\begin{table*}[h]

\begin{center}
\flushleft {\bf Table 3.} (Continued)

\begin{tabular}{rlllrrcll} \\[-0.22cm] \hline \\[-0.35cm]
\multicolumn{1}{c}{\#}               &
\multicolumn{1}{c}{Name}             &
\multicolumn{1}{c}{$\alpha\,(1950)$} &
\multicolumn{1}{c}{$\delta\,(1950)$} &
\multicolumn{1}{c}{$V_{hel}^{\,a}\pm\sigma_V$}    &
\multicolumn{1}{r}{m$_{B}$}          &
\multicolumn{1}{c}{M$_{B}^{\;\;b}$}  &
\multicolumn{1}{c}{Type}             &
\multicolumn{1}{l}{Other names from NED}\\
&&&&&&&& \multicolumn{1}{c}{and number of reference}\\
&
\multicolumn{1}{c}{ (1) } &
\multicolumn{1}{c}{ (2) } &
\multicolumn{1}{c}{ (3) } &
\multicolumn{1}{c}{ (4) } &
\multicolumn{1}{c}{ (5) } &
\multicolumn{1}{c}{ (6) } &
\multicolumn{1}{c}{ (7) } &
\multicolumn{1}{c}{ (8) } \\
\hline
\qq& HS 1324+4019            &13 24 05.3 &+40 19 39  &7671  $\pm$ 39    &17.60 &--17.45&BCG      & \\
\qq& HS 1331+3657            &13 31 46.5 &+36 57 21  &17949 $\pm$ 22    &18.00 &--18.90&BCG      & {\bf 3}\\
\qq& HS 1344+3511            &13 44 53.9 &+35 11 21  &16101 $\pm$ 57    &16.80 &--19.86&SBN      &CG 1189, {\bf 2} \\
\qq& HS 1347+3553            &13 47 12.2 &+35 53 48  &17062 $\pm$ 111   &19.50 &--17.28&DANS?    & \\
\qq& HS 1347+3811            &13 47 02.3 &+38 11 35  &2804  $\pm$ 36    &18.80 &--14.06&BCG      & {\bf 3}, {\bf 6}\\
\qq& HS 1349+3942            &13 49 22.9 &+39 42 04  &1752  $\pm$ 50    &16.70 &--15.14&BCG      & {\bf 5}, {\bf 6}\\
\qq& HS 1350+3538$^\dagger$  &13 50 13.8 &+35 38 37  &17791 $\pm$ 519   &17.90 &--18.98&SBN?     & \\
\qq& HS 1400+3927            &14 00 29.9 &+39 27 38  &1396  $\pm$ 39    &17.20 &--14.15&BCG      &CG 330, {\bf 9} \\
\qq& HS 1408+3857            &14 08 37.6 &+38 57 11  &7701  $\pm$ 37    &16.80 &--18.26&BCG      &CG 369, {\bf 9} \\
\qq& HS 1425+3835$^\ddagger$ &14 25 14.6 &+38 35 33  &6548  $\pm$ 21    &16.80 &--17.91&BCG?     &CG 0435, {\bf 5}, {\bf 6} \\
\qq& HS 1435+3645$^\ddagger$ &14 35 23.5 &+36 45 12  &15655 $\pm$ 48    &+17.10&--19.50&BCG?     & \\
\qq& HS 1440+3805            &14 40 08.8 &+38 05 06  &9599  $\pm$ 27    &16.70 &--18.84&DANS?    &NPM1G +38.0321, {\bf 5}, {\bf 6} \\
\qq& HS 1440+3828$^\ddagger$ &14 40 40.0 &+38 28 52  &20761 $\pm$ 12    &17.28 &--19.93&SBN?     & \\
\qq& HS 1450+3903$^\ddagger$ &14 50 04.8 &+39 03 11  &18096 $\pm$ 42    &+16.83&--20.08&SBN      & \\
\qq& HS 1502+4152            &15 02 31.3 &+41 52 35  &5004  $\pm$ 30    &+16.35&--17.77&BCG      & {\bf 5}, {\bf 6} \\
\qq& HS 1504+3922            &15 04 15.6 &+39 22 16  &8962  $\pm$ 54    &18.80 &--16.59&BCG      &CG 624, {\bf 9} \\
\qq& HS 1504+3923            &15 04 10.9 &+39 23 21  &9189  $\pm$ 36    &18.30 &--17.14&BCG      & CG 623 \\
\qq& HS 1509+4409            &15 09 56.8 &+44 09 28  &8284  $\pm$ 69    &      &       &SBN?     & \\
\qq& HS 1519+3948$^\ddagger$ &15 19 49.0 &+39 48 19  &14716 $\pm$ 21    &17.60 &--18.86&BCG?     & \\
\qq& HS 1520+3717            &15 20 30.6 &+37 17 56  &11239 $\pm$ 63    &+15.42&--20.46&BCG      & {\bf 3}\\
\qq& HS 1521+3617$^\ddagger$ &15 21 04.9 &+36 17 24  &6998  $\pm$ 24    &+16.88&--17.97&BCG      & \\
\qq& HS 1524+4205             &15 24 08.0 &+42 05 01  &6700  $\pm$ 30    &18.40 &--16.36&DANS?    & {\bf 5}\\
\qq& HS 1526+4045             &15 26 56.7 &+40 45 17  &8610  $\pm$ 9     &17.70 &--17.60&DANS?    & {\bf 5}\\
\qq& HS 1528+3858$^\ddagger$  &15 28 43.3 &+38 58 11  &10579 $\pm$ 69    &17.60 &--18.15&DANS?    & \\
\qq& HS 1543+3602             &15 43 12.5 &+36 02 46  &1889  $\pm$ 18    &16.00 &--16.00&BCG      & ABELL 2124 \\
\qq& HS 1543+4525             &15 43 23.3 &+45 25 45  &11519 $\pm$ 27    &17.40 &--18.53&DANS?    & {\bf 5}, {\bf 6}\\
\qq& HS 1544+3803$^\ddagger$  &15 44 21.6 &+38 03 12  &12008 $\pm$ 27    &17.20 &--18.82&BCG?     & \\
\qq& HS 1546+3526             &15 46 54.3 &+35 26 24  &16525 $\pm$ 27    &18.20 &--18.52&BCG      & \\
\qq& HS 1546+4755             &15 46 56.5 &+47 55 34  &11250 $\pm$ 9     &18.90 &--16.98&BCG      & {\bf 5}, {\bf 6}\\
\qq& HS 1558+3543             &15 58 27.3 &+35 43 15  &20284 $\pm$ 69    &18.30 &--18.87&BCG      & {\bf 3}\\
\qq& HS 1608+3654             &16 08 31.5 &+36 54 42  &9586  $\pm$ 12    &17.30 &--18.23&BCG      & {\bf 3}\\
\qq& HS 1609+4827             &16 09 44.4 &+48 27 44  &2817  $\pm$ 24    &16.40 &--16.47&BCG      & {\bf 5}, {\bf 6}\\
\qq& HS 1612+3650             &16 12 34.7 &+36 50 06  &8750  $\pm$ 24    &18.60 &--16.73&BCG      & \\
\qq& HS 1612+3720$^\ddagger$  &16 12 30.0 &+37 20 35  &11296 $\pm$ 12    &17.50 &--18.39&BCG      & \\
\qq& HS 1617+3757             &16 17 32.0 &+37 57 55  &9118  $\pm$ 15    &18.20 &--17.22&BCG      & \\
\qq& HS 1619+3523             &16 19 47.3 &+35 23 16  &15274 $\pm$ 66    &16.80 &--19.74&LINER?   & KUG 1619+353B \\
\qq& HS 1619+3752             &16 19 55.7 &+37 52 35  &9883  $\pm$ 9     &17.30 &--18.30&BCG      & PC 1619+3752, {\bf 8} \\
\qq& HS 1620+4003$^\dagger$   &16 20 49.9 &+40 03 47  &18903 $\pm$ 45    &16.20 &--20.81&LINER    & \\
\qq& HS 1627+3927$^\ddagger$  &16 27 27.4 &+39 27 26  &7984  $\pm$ 9     &17.80 &--17.34&BCG      &ABELL 2199:[BO85]156 \\
\qq& HS 1627+3945$^\ddagger$  &16 27 04.8 &+39 45 29  &8122  $\pm$ 15    &+15.11&--20.06&SBN?     &ABELL 2199:[BO85]107 \\
\hline \\[-0.2cm]
\multicolumn{9}{l}{ $^a$ Heliocentric velocities.; $^b$ Absolute magnitudes are not corrected for galactic extinction}\\
\multicolumn{9}{l}{ + APM magnitudes; $^\dagger$ objects from the  second priority random sample; $^\ddagger$ galaxies from the APM selected sample} \\
\multicolumn{9}{l}{{\bf References:}
{\bf 1} -- Augarde et al. (\cite{Augarde94}); {\bf 2} -- de Grijp et al. (\cite{Grijp92}); {\bf 3} -- Hopp et al. (\cite{PaperIII}); {\bf 4} -- Huchra et al. (\cite{Huchra95})} \\
\multicolumn{9}{l}{{\bf 5} -- Popescu et al. (\cite{Popescu96}); {\bf 6} -- Popescu et al. (\cite{Popescu99}); {\bf 7} -- Ramella et al. (\cite{Ramella95}); {\bf 8} -- Schneider et al. (\cite{SCHNEIDER94})} \\
\multicolumn{9}{l}{{\bf 9} -- Tift et al. (\cite{Tift86}); {\bf 10} -- Weistrop \& Downes (\cite{Weistrop91})}\\
\end{tabular}
\end{center}
\end{table*}




\scriptsize
\setcounter{qub}{0}

\begin{table*}[h]

\begin{center}
\caption{\label{Tab3} {\bf Emission line parameters}}

\begin{tabular}{rlrcccccrrr} \hline \\[-0.35cm]
\multicolumn{1}{c}{\#}                             &
\multicolumn{1}{c}{ Name}                          &
\multicolumn{1}{c}{ F(H$\beta$)$^a$ }              &
\multicolumn{1}{c}{ $\frac{F(\lambda 3727)}{F(H\beta)}$} &
\multicolumn{1}{c}{ $\frac{F(\lambda 5007)}{F(H\beta)}$} &
\multicolumn{1}{c}{ $\frac{F(H\alpha)}{F(H\beta)}$} &
\multicolumn{1}{c}{ $\frac{F(\lambda 6583)}{F(H\alpha)}$}&
\multicolumn{1}{c}{ $\frac{F([SII])}{F(H\alpha)}$} &
\multicolumn{1}{c}{ W$_{\lambda 3727}$(\AA)}       &
\multicolumn{1}{c}{ W$_{H\beta}$(\AA)}             &
\multicolumn{1}{c}{ W$_{\lambda 5007}$(\AA)}       \\
&
\multicolumn{1}{c}{ (1) } &
\multicolumn{1}{c}{ (2) } &
\multicolumn{1}{c}{ (3) } &
\multicolumn{1}{c}{ (4) } &
\multicolumn{1}{c}{ (5) } &
\multicolumn{1}{c}{ (6) } &
\multicolumn{1}{c}{ (7) } &
\multicolumn{1}{c}{ (8) } &
\multicolumn{1}{c}{ (9) } &
\multicolumn{1}{c}{ (10) } \\
\hline
\qq& HS 0732+3529            &45 & 2.53 &  10.09&   5.89&   0.03&    --  &66   &21   &237  \\
\qq& HS 0758+3901            &15 & 4.93 &   6.00&   4.40&   0.06&   0.23 &108  &28   &199  \\
\qq& HS 0759+3811            &37 & 2.22 &   2.86&   3.38&   0.06&   0.15 &31   &18   &56   \\
\qq& HS 0801+3740            &42 & 3.17 &   2.55&   3.38&   0.11&   0.35 &96   &38   &106  \\
\qq& HS 0801+3746$^\dagger$  &+47& --   &    -- &    -- &    -- &    --  &--   &+12  &--   \\
\qq& HS 0803+3648            &33 & 3.21 &   3.52&   3.64&   0.09&   0.23 &65   &29   &111  \\
\qq& HS 0826+3521            &+56& --   &    -- &    -- &    -- &   0.50 &--   &+38  &--   \\
\qq& HS 0834+3741            &36 & 2.78 &   2.33&   1.14&    -- &    --  &85   &60   &127  \\
\qq& HS 0837+3816            &13 & 6.92 &   4.77&   4.38&   0.35&    --  &66   &10   &48   \\
\qq& HS 0840+3938            &42 & 4.45 &   3.17&   3.43&   0.07&   0.36 &67   &16   &53   \\
\qq& HS 0840+4013            &+61& --   &    -- &    -- &   0.49&    --  &67   &+36  &--   \\
\qq& HS 0841+3959            &7  & 8.29 &   1.57&   7.29&   0.37&    --  &122  &6    &10   \\
\qq& HS 0843+3637            &326& 2.64 &   4.28&   4.47&   0.04&   0.17 &112  &66   &343  \\
\qq& HS 0847+3639            &87 & 4.36 &   1.64&   3.92&   0.11&   0.22 &56   &14   &23   \\
\qq& HS 0852+3614            &5  & 6.00 &   4.40&   4.00&    -- &    --  &42   &7    &30   \\
\qq& HS 0852+4003            &81 & 2.86 &   4.94&   3.77&    -- &   0.16 &52   &23   &118  \\
\qq& HS 0900+3651            &38 & 2.16 &   4.63&   3.37&   0.05&   0.23 &81   &52   &225  \\
\qq& HS 0902+3759            &12 & 5.58 &   2.00&  12.42&   0.42&   0.54 &41   &6    &11   \\
\qq& HS 0902+4005            &82 & 3.54 &   3.83&   3.26&   0.04&   0.25 &94   &32   &130  \\
\qq& HS 0903+4349*           &7: & --   &   6.00&    -- &    -- &    --  &--   &28   &196  \\
\qq& HS 0904+4357*           &5: & 3.40 &   1.80&    -- &    -- &    --  &42   &16   &25   \\
\qq& HS 0904+4400*           &10:& 1.50 &   2.40&    -- &    -- &    --  &43   &32   &67   \\
\qq& HS 0905+3948            &44 & 3.05 &   2.77&   3.98&    -- &   0.15 &56   &20   &57   \\
\qq& HS 0907+3728            &111& 1.53 &   0.52&   5.26&   0.38&   0.25 &24   &14   &7    \\
\qq& HS 0913+3516            &19 & 3.79 &   2.95&   2.89&   0.09&   0.51 &94   &17   &51   \\
\qq& HS 0925+3529            &44 & 2.30 &   4.55&   3.39&   0.01&   0.18 &59   &24   &115  \\
\qq& HS 0928+3725$^\dagger$  &64 & 3.94 &   1.63&   3.41&   0.17&   0.34 &68   &20   &33   \\
\qq& HS 0928+4006            &40 & 1.35 &   5.20&   3.50&   0.04&   0.19 &39   &21   &109  \\
\qq& HS 0930+3548            &25 & 4.32 &   3.40&   4.04&   0.09&   0.10 &51   &10   &36   \\
\qq& HS 0934+3629$^\dagger$  &88 & --   &   0.74&   4.77&   0.18&    --  &--   &33   &25   \\
\qq& HS 0936+3601$^\dagger$  &5  & --   &   2.80&   4.80&    -- &    --  &--   &7    &18   \\
\qq& HS 0936+3648$^\dagger$  &47 & 3.32 &   0.89&   3.49&   0.24&   0.30 &28   &9    &8    \\
\qq& HS 0938+3544$^\dagger$  &+23& --   & +0.78 &    -- &    -- &    --  &--   &+108 &45   \\
\qq& HS 0942+3600            &26 & 2.50 &   2.73&   4.62&   0.13&   0.39 &98   &23   &72   \\
\qq& HS 0942+3644            &30 & 3.73 &   2.00&   4.57&   0.24&   0.43 &39   &9    &17   \\
\qq& HS 0947+3559            &30 & 2.50 &   1.50&   4.47&   0.18&   0.31 &43   &15   &21   \\
\qq& HS 0951+3606            &24 & 1.50 &   2.33&   6.67&   0.21&   0.34 &51   &16   &37   \\
\qq& HS 0951+3841            &42 & 3.76 &   3.19&   4.55&   0.08&   0.28 &122  &34   &115  \\
\qq& HS 0952+3803            &4  & --   &   2.75&   2.75&    -- &    --  &--   &10   &27   \\
\qq& HS 0958+4043*           &19:& 2.05 &   5.95&    -- &    -- &    --  &105  &73   &283  \\
\hline  \\[--0.2cm]
\end{tabular}
\end{center}
\end{table*}

\newpage

\begin{table*}[h]

\begin{center}
\flushleft {\bf Table 4.} (Continued)
\begin{tabular}{rlrcccccrrr} \hline \\[-0.35cm]
\multicolumn{1}{c}{\#}                             &
\multicolumn{1}{c}{ Name}                          &
\multicolumn{1}{c}{ F(H$\beta$)$^a$ }              &
\multicolumn{1}{c}{ $\frac{F(\lambda 3727)}{F(H\beta)}$} &
\multicolumn{1}{c}{ $\frac{F(\lambda 5007)}{F(H\beta)}$} &
\multicolumn{1}{c}{ $\frac{F(H\alpha)}{F(H\beta)}$} &
\multicolumn{1}{c}{ $\frac{F(\lambda 6583)}{F(H\alpha)}$}&
\multicolumn{1}{c}{ $\frac{F([SII])}{F(H\alpha)}$} &
\multicolumn{1}{c}{ W$_{\lambda 3727}$(\AA)}       &
\multicolumn{1}{c}{ W$_{H\beta}$(\AA)}             &
\multicolumn{1}{c}{ W$_{\lambda 5007}$(\AA)}       \\
&
\multicolumn{1}{c}{ (1) } &
\multicolumn{1}{c}{ (2) } &
\multicolumn{1}{c}{ (3) } &
\multicolumn{1}{c}{ (4) } &
\multicolumn{1}{c}{ (5) } &
\multicolumn{1}{c}{ (6) } &
\multicolumn{1}{c}{ (7) } &
\multicolumn{1}{c}{ (8) } &
\multicolumn{1}{c}{ (9) } &
\multicolumn{1}{c}{ (10) } \\
\hline
\qq& HS 1014+4030*           &69:& 2.52 &   3.25&    -- &    -- &    --  &102  &39   &139  \\
\qq& HS 1014+4916*           &10:& 4.90 &   1.00&    -- &    -- &    --  &102  &15   &16   \\
\qq& HS 1018+3847            &21 & 4.10 &   4.38&   6.24&   0.11&   0.27 &24   &5    &21   \\
\qq& HS 1021+3637$^\dagger$  &+18& --   &  +0.67&    -- &    -- &    --  &--   &+34  &19   \\
\qq& HS 1021+4218*           &13:& 2.46 &   3.92&    -- &    -- &    --  &51   &91   &211  \\
\qq& HS 1025+4005$^\dagger$  &72 & 3.07 &   3.22&   4.04&   0.09&   0.25 &54   &19   &61   \\
\qq& HS 1026+3521            &15 & 1.87 &   7.53&   5.60&   0.05&   0.25 &49   &20   &166  \\
\qq& HS 1028+3843            &135& 0.44 &   7.03&   2.99&   0.00&    --  &70   &404  &2819 \\
\qq& HS 1047+3714$^\dagger$  &+38& --   &  +2.24&    -- &   1.97&   0.87 &14   &+6   &15   \\
\qq& HS 1048+4026            &+46& --   &  +0.78&    -- &    -- &    --  &14   &+43  &33   \\
\qq& HS 1050+4002             &30 & 3.90 &   3.87&   3.47&   0.02&    --  &106  &31   &122  \\
\qq& HS 1059+3934             &25 & 2.12 &   5.52&   3.68&    -- &    --  &101  &56   &335  \\
\qq& HS 1102+3644             &+31& --   &    -- &    -- &   0.58&    --  &--   &+11  &--   \\
\qq& HS 1107+3524             &20 & 4.55 &   3.90&   4.35&   0.07&   0.23 &75   &12   &47   \\
\qq& HS 1107+3637             &21 & 2.43 &   5.24&   3.67&   0.04&   0.12 &--   &39   &210  \\
\qq& HS 1107+3831             &63 & 3.60 &   3.71&   4.73&   0.14&   0.26 &42   &12   &50   \\
\qq& HS 1116+3951             &34 & 5.21 &   1.62&   3.00:&  0.26&   0.35 &107  &24   &44   \\
\qq& HS 1118+3842             &13 & 4.54 &   6.15&   3.00&   0.23&    --  &118  &27   &170  \\
\qq& HS 1124+3635             &102& 1.15 &   6.17&   3.10&    -- &   0.09 &98   &111  &810  \\
\qq& HS 1125+3624             &41 & 3.41 &   3.39&   3.46&   0.14&   0.27 &127  &30   &112  \\
\qq& HS 1125+3748             &17 & 5.88 &   5.29&   4.94&   0.07&   0.23 &77   &10   &56   \\
\qq& HS 1126+3701             &32 & 3.38 &   6.09&   3.53&   0.06&   0.25 &44   &12   &78   \\
\qq& HS 1129+3730             &40 & 3.05 &   3.85&   3.70&    -- &    --  &77   &35   &137  \\
\qq& HS 1134+3639             &10 & 7.30 &   5.80&   4.70&   0.09&   0.36 &48   &8    &55   \\
\qq& HS 1134+3640             &103& 2.15 &   4.22&   2.82&   0.03&   0.14 &97   &146  &689  \\
\qq& HS 1135+3709             &+7 & --   &  +0.71&    -- &    -- &    --  &--   &+19  &9    \\
\qq& HS 1144+3552             &23 & 2.96 &   4.74&   3.26&    -- &    --  &201  &151  &686  \\
\qq& HS 1145+3709             &32 & 3.69 &   2.06&   4.63&   0.24&   0.40 &56   &14   &30   \\
\qq& HS 1147+3653             &+18& --   &  +0.33&    -- &   0.33&   0.72 &--   &+37  &10   \\
\qq& HS 1150+3756             &14 & 3.29 &   3.29&   2.21&    -- &    --  &141  &26   &90   \\
\qq& HS 1159+3701             &16 & 3.19 &   3.38&   4.25&   0.12&   0.25 &58   &19   &61   \\
\qq& HS 1221+3721             &27 & 1.30 &   4.78&   2.70&   0.03&   0.21 &41   &41   &190  \\
\qq& HS 1228+3632             &34 & 2.41 &   2.65&   2.50&   0.15&   0.35 &163  &47   &123  \\
\qq& HS 1240+3902             &77 & 1.12 &   5.36&   3.12&    -- &   0.08 &83   &101  &498  \\
\qq& HS 1300+3646             &13 & 1.85 &   6.31&   4.69&   0.05&    --  &70   &37   &300  \\
\qq& HS 1301+3835             &6  & 8.00 &   5.67&   6.50&   0.21&   0.72 &21   &3    &20   \\
\qq& HS 1308+3545             &8  & --   &   4.38&   4.25&    -- &    --  &--   &200: &333: \\
\qq& HS 1309+3806             &50 & 0.48 &   8.30&   3.96&    -- &    --  &73   &218  &1417 \\
\qq& HS 1318+3624             &26 & 2.54 &   4.62&   4.08&   0.07&   0.25 &81   &31   &138  \\
\qq& HS 1324+4019             &21 & 3.14 &   4.57&   4.19&   0.09&   0.31 &106  &24   &114  \\
\hline  \\[--0.2cm]
\end{tabular}
\end{center}
\end{table*}

\newpage

\begin{table*}[h]

\begin{center}
\flushleft {\bf Table 4.} (Continued)
\begin{tabular}{rlrcccccrrr} \hline \\[-0.35cm]
\multicolumn{1}{c}{\#}                             &
\multicolumn{1}{c}{ Name}                          &
\multicolumn{1}{c}{ F(H$\beta$)$^a$ }              &
\multicolumn{1}{c}{ $\frac{F(\lambda 3727)}{F(H\beta)}$} &
\multicolumn{1}{c}{ $\frac{F(\lambda 5007)}{F(H\beta)}$} &
\multicolumn{1}{c}{ $\frac{F(H\alpha)}{F(H\beta)}$} &
\multicolumn{1}{c}{ $\frac{F(\lambda 6583)}{F(H\alpha)}$}&
\multicolumn{1}{c}{ $\frac{F([SII])}{F(H\alpha)}$} &
\multicolumn{1}{c}{ W$_{\lambda 3727}$(\AA)}       &
\multicolumn{1}{c}{ W$_{H\beta}$(\AA)}             &
\multicolumn{1}{c}{ W$_{\lambda 5007}$(\AA)}       \\
&
\multicolumn{1}{c}{ (1) } &
\multicolumn{1}{c}{ (2) } &
\multicolumn{1}{c}{ (3) } &
\multicolumn{1}{c}{ (4) } &
\multicolumn{1}{c}{ (5) } &
\multicolumn{1}{c}{ (6) } &
\multicolumn{1}{c}{ (7) } &
\multicolumn{1}{c}{ (8) } &
\multicolumn{1}{c}{ (9) } &
\multicolumn{1}{c}{ (10) } \\
\hline
\qq& HS 1331+3657*            &87:& 1.16 &   6.55&    -- &    -- &    --  &49   &81   &525  \\
\qq& HS 1344+3511             &126& 2.06 &   2.81&   5.56&   0.24&   0.20 &111  &65   &184  \\
\qq& HS 1347+3553             &13 & 1.69 &    -- &   2.77&    -- &    --  &27   &16   &--   \\
\qq& HS 1347+3811             &10 & 1.30 &   4.70&   4.60&   0.04&   0.17 &27   &22   &108  \\
\qq& HS 1349+3942             &23 & 2.43 &   3.26&   5.30&   0.11&   0.29 &41   &15   &54   \\
\qq& HS 1350+3538$^\dagger$   &+19& --   &  +0.32&    -- &    -- &    --  &--   &+15  &6    \\
\qq& HS 1400+3927             &107& 1.55 &   5.39&   3.44&   0.03&   0.14 &94   &66   &375  \\
\qq& HS 1408+3857             &53 & 3.15 &   3.91&   3.64&   0.05&   0.15 &57   &19   &76   \\
\qq& HS 1425+3835$^\ddagger$  &45 & 4.20 &   2.11&    -- &    -- &    --  &94   &11   &25   \\
\qq& HS 1435+3645$^\ddagger$  &9  & --   &   3.11&   4.33&   0.13&   0.41 &--   &11   &26   \\
\qq& HS 1440+3805             &14 & 2.14 &   0.57&    -- &    -- &    --  &40   &10   &5    \\
\qq& HS 1440+3828$^\ddagger$  &+16& --   &    -- &    -- &   0.31&    --  &44   &+53  &--   \\
\qq& HS 1450+3903$^\ddagger$  &25 & 2.72 &   0.64&   4.08&   0.34&   0.22 &39   &11   &7    \\
\qq& HS 1502+4152             &22 & 2.09 &   3.55&    -- &    -- &    --  &50   &34   &109  \\
\qq& HS 1504+3922             &97 & 0.91 &   5.69&   3.04&   0.02&   0.13 &83   &151  &854  \\
\qq& HS 1504+3923             &68 & 1.29 &   5.96&   3.41&    -- &    --  &65   &66   &402  \\
\qq& HS 1509+4409             &7  & 8.43 &   3.86&    -- &    -- &    --  &31   &3    &11   \\
\qq& HS 1519+3948$^\ddagger$  &6  & 12.50&   5.67&   3.67&   0.27&    --  &114  &12   &53   \\
\qq& HS 1520+3717             &11 & 5.36 &   4.73&   4.36&   0.10&   0.46 &81   &7    &36   \\
\qq& HS 1521+3617$^\ddagger$  &18 & 2.39 &   4.67&   3.39&   0.07&    --  &61   &16   &81   \\
\qq& HS 1524+4205             &5  & 7.20 &   2.20&    -- &    -- &    --  &58   &3    &7    \\
\qq& HS 1526+4045             &4  & 8.00 &   3.00&    -- &    -- &    --  &23   &2    &6    \\
\qq& HS 1528+3858$^\ddagger$  &+14& --   &  +0.93&    -- &    -- &    --  &--   &+58  &35   \\
\qq& HS 1543+3602             &226& 1.82 &   5.14&   3.17&    -- &   0.10 &88   &46   &246  \\
\qq& HS 1543+4525             &32 & 3.78 &   1.22&   3.69&   0.25&   0.39 &54   &14   &17   \\
\qq& HS 1544+3803$^\ddagger$  &33 & 3.33 &   2.61&   4.67&   0.23&   0.20 &132  &14   &39   \\
\qq& HS 1546+3526             &33 & 2.30 &   4.15&   3.18&   0.06&   0.27 &110  &46   &194  \\
\qq& HS 1546+4755             &105& 2.86 &   5.08&   3.31&   0.07&   0.09 &152  &55   &261  \\
\qq& HS 1558+3543             &40 & 2.92 &   3.80&   3.00&   0.06&   0.12 &234  &58   &228  \\
\qq& HS 1608+3654             &23 & 6.35 &   4.04&   4.65&   0.04&   0.32 &191  &11   &46   \\
\qq& HS 1609+4827             &70 & 2.79 &   3.33&   4.31&   0.09&   0.27 &34   &13   &44   \\
\qq& HS 1612+3650             &38 & 2.79 &   4.08&   3.11&   0.03&   0.19 &103  &31   &133  \\
\qq& HS 1612+3720$^\ddagger$  &17 & 6.41 &   3.94&   5.24&   0.12&   0.28 &86   &10   &41   \\
\qq& HS 1617+3757             &149& 0.98 &   6.68&   2.93&    -- &   0.07 &171  &215  &1320 \\
\qq& HS 1619+3523             &52 & 1.87 &   3.85&   4.48&   0.60&   0.38 &25   &9    &35   \\
\qq& HS 1619+3752             &55 & 2.36 &   2.93&   3.69&   0.10&   0.29 &47   &27   &76   \\
\qq& HS 1620+4003$^\dagger$   &70 & 1.94 &   5.90&   7.06&   0.56&   0.46 &22   &10   &60   \\
\qq& HS 1627+3927$^\ddagger$  &33 & 1.55 &   4.55&   3.24&   0.05&   0.14 &64   &50   &233  \\
\qq& HS 1627+3945$^\ddagger$  &15 & 2.73 &   3.00&   8.07&   0.21&   0.37 &24   &5    &14   \\
\hline  \\[-0.2cm]
\multicolumn{10}{l}{ $^a$ Flux in 10$^{-16}$ ergs s$^{-1}$ sm$^{-2}$ \AA$^{-1}$; * objects observed during non-photometric conditions} \\
\multicolumn{10}{l}{$^\dagger$ objects from the second priority random sample; $^\ddagger$ galaxies from the APM selected sample}  \\
\multicolumn{10}{l}{ $^+$ Parameters for H$\alpha$ emission line; : Parameters with less confident values} \\
\end{tabular}
\end{center}
\end{table*}




\setcounter{qub}{0}

\begin{table*}[h]

\begin{center}
\caption{\label{Tab4} Coordinates, Redshifts and Magnitudes of QSOs}

\begin{tabular}{rlccrrl} \hline \\[-0.35cm]
\MC{1}{c}{\#}               &
\MC{1}{c}{Name}             &
\MC{1}{c}{$\alpha\,(1950)$} &
\MC{1}{c}{$\delta\,(1950)$} &
\MC{1}{c}{$z^{\,a}$}        &
\MC{1}{c}{$m_{B}$}          &
\MC{1}{c}{Detected {\bf emission lines} }\\
&
\MC{1}{c}{ (1) } &
\MC{1}{c}{ (2) } &
\MC{1}{c}{ (3) } &
\MC{1}{c}{ (4) } &
\MC{1}{c}{ (5) } &
\MC{1}{c}{ (6) }
\\ \hline
\qq & HS 0833+3516          &08 33 50.4 &+35 16 18  &3.304 &19.00 & Ly$\alpha$ 1216 \AA, Si{\sc iv}/O{\sc iv}] 1400 \AA, C{\sc iv} 1549 \AA  \\
\qq & HS 0844+3842          &08 44 01.4 &+38 42 13  &3.177 &19.70 & Ly$\alpha$ 1216 \AA, Si{\sc iv}/O{\sc iv}] 1400 \AA, C{\sc iv} 1549 \AA \\
\qq & HS 0855+3724          &08 55 59.7 &+37 24 42  &3.207 &19.60 & Ly$\alpha$ 1216 \AA, C{\sc iv} 1549 \AA \\
\qq & HS 0857+3601$^\dagger$&08 57 55.0 &+36 01 15  &3.153 &17.80 & Ly$\alpha$ 1216 \AA, Si{\sc iv}/O{\sc iv}] 1400 \AA, C{\sc iv} 1549 \AA \\
\qq & HS 0954+3549          &09 54 36.8 &+35 49 40  &3.296 &+19.20& Ly$\alpha$ 1216 \AA, Si{\sc iv}/O{\sc iv}] 1400 \AA, C{\sc iv} 1549 \AA \\
\qq & HS 1005+3638          &10 05 44.2 &+36 38 01  &3.162 &17.50 & Ly$\alpha$ 1216 \AA, Si{\sc iv}/O{\sc iv}] 1400 \AA, C{\sc iv} 1549 \AA \\
\qq & HS 1143+3954          &11 43 10.9 &+39 54 26  &3.189 &19.40 & Ly$\alpha$ 1216 \AA, Si{\sc iv}/O{\sc iv}] 1400 \AA, C{\sc iv} 1549 \AA \\
\qq & HS 1215+3821          &12 15 13.1 &+38 21 45  &3.161 &+20.70& Ly$\alpha$ 1216 \AA, Si{\sc iv}/O{\sc iv}] 1400 \AA, C{\sc iv} 1549 \AA \\
\hline \\[-0.2cm]
\multicolumn{7}{l}{ $^a$ Observed redshift; $^+$ APM magnitudes; $^\dagger$ objects from the second priority random sample} \\
\end{tabular}
\end{center}
\end{table*}




\setcounter{qub}{0}
\begin{table*}[h]

\begin{center}
\caption{\label{Tab5} Galaxies without detected emission lines }

\begin{tabular}{rlllrccl} \hline \\[-0.35cm]
\multicolumn{1}{c}{\#}                             &
\multicolumn{1}{c}{Name} &
\multicolumn{1}{c}{$\alpha\,(1950)$} &
\multicolumn{1}{c}{$\delta\,(1950)$} &
\multicolumn{1}{c}{$v^{\,a}_{0}$} &
\multicolumn{1}{c}{m$_B$} &
\multicolumn{1}{r}{M$_B^{\,b}$} &
\multicolumn{1}{l}{Absorption lines} \\
&
\MC{1}{c}{ (1) } &
\MC{1}{c}{ (2) } &
\MC{1}{c}{ (3) } &
\MC{1}{c}{ (4) } &
\MC{1}{c}{ (5) } &
\MC{1}{c}{ (6) } &
\MC{1}{c}{ (7) }
\\ \hline
\qq& HS 0812+3516$^\dagger$ & 08 12 23.1 &+35 16 23  &33466 &18.40 &--19.85 & CaK, CaH, H$\beta$, NaD   \\
\qq& HS 0843+3536           & 08 43 49.9 &+35 36 24  &15883 &18.80 &--17.83 & CaK, CaH, H$\beta$, NaD, H$\alpha$   \\
\qq& HS 0929+3956           & 09 29 11.9 &+39 56 52  &6031  &+19.27&--15.26 & CaK, CaH, G$_{band}$, NaD, H$\alpha$   \\
\qq& HS 0936+3753           & 09 36 14.4 &+37 53 25  &46285:&18.00 &--20.95 & CaK, CaH, G$_{band}$, H$\beta$, MgIb, H$\alpha$   \\

\hline \\[-0.2cm]
\multicolumn{8}{l}{ $^a$ Heliocentric velocities; $^b$ Absolute magnitudes are not corrected for galactic extinction} \\
\multicolumn{8}{l}{ $^\dagger$ objects from the second priority random sample; $^+$ APM magnitude; : less confident values} \\

\end{tabular}
\end{center}

\end{table*}




\setcounter{qub}{0}

\begin{table*}[t]
\begin{center}
\caption{\label{Tab6}Stars}
\begin{tabular}{rlccrcl} \hline \\[-0.35cm]
\multicolumn{1}{c}{\#}             &
\multicolumn{1}{c}{Name}             &
\multicolumn{1}{c}{$\alpha\,(1950)$} &
\multicolumn{1}{c}{$\delta\,(1950)$} &
\multicolumn{1}{r}{m$_{B}$}          &
\multicolumn{1}{l}{Type}             &
\multicolumn{1}{c}{Spectral features} \\
&
\MC{1}{c}{ (1) } &
\MC{1}{c}{ (2) } &
\MC{1}{c}{ (3) } &
\MC{1}{c}{ (4) } &
\MC{1}{c}{ (5) } &
\MC{1}{c}{ (6) }
\\ \hline
\qq &HS 0737+3902           & 07 37 24.4 & +39 02 30  & 17.40 & G  & CaK, CaH, G$_{band}$, MgIb, CaFe5269, H$_{\alpha}$ \\
\qq &HS 0745+3600           & 07 45 35.6 & +36 00 22  & 18.80 &A-F & H$_{\beta}$, H$_{\alpha}$ \\
\qq &HS 0750+3637           & 07 50 51.5 & +36 37 07  & 18.30 & G  & CaK, CaH, G$_{band}$, MgIb, NaD, H$_{\alpha}$ \\
\qq &HS 0753+3519$^\dagger$ & 07 53 09.6 & +35 19 59  & 16.70 & F  & CaK, CaH, H$_{\delta}$, H$_{\gamma}$, H$_{\beta}$, CaFe5269, H$_{\alpha}$ \\
\qq &HS 0802+3809           & 08 02 01.0 & +38 09 06  & 18.30 &A-F & H$_{\beta}$, H$_{\alpha}$ \\
\qq &HS 0840+3615           & 08 40 50.4 & +36 15 09  & 17.10 & F  & CaK, CaH, H$_{\beta}$, H$_{\alpha}$ \\
\qq &HS 0859+3519$^\dagger$ & 08 59 35.5 & +35 19 21  & 19.00 & M  &  \\
\qq &HS 0906+3846$^\dagger$ & 09 06 55.1 & +38 46 29  & 18.50 & G  & CaK, CaH, G$_{band}$, H$_{\beta}$, MgIb, CaFe5269 \\
\qq &HS 0929+3657$^\dagger$ & 09 29 38.9 & +36 57 18  & 19.40 &F-G & CaK, CaH, H$_{\beta}$, MgIb, CaFe5269, H$_{\alpha}$ \\
\qq &HS 0940+3753$^\dagger$ & 09 40 19.1 & +37 53 33  & 17.10 & F  & H$_{\beta}$, MgIb, H$_{\alpha}$ \\
\qq &HS 0941+3659$^\dagger$ & 09 41 52.0 & +36 59 20  & 16.50 & F  & CaK, CaH, G$_{band}$, H$_{\beta}$, MgIb, H$_{\alpha}$  \\
\qq &HS 1031+3524$^\dagger$ & 10 31 37.2 & +35 24 12  & 17.00 & A  & H$_{\gamma}$, H$_{\beta}$, H$_{\alpha}$  \\
\qq &HS 1132+3904           & 11 32 17.3 & +39 04 43  &+17.84 & A  & H$_{\gamma}$, H$_{\beta}$, H$_{\alpha}$ \\
\qq &HS 1245+3842$^\dagger$ & 12 45 45.7 & +38 42 19  & 16.80 & F  & CaK, CaH, H$_{\beta}$, MgIb, NaD, H$_{\alpha}$ \\
\qq &HS 1300+3748$^\dagger$ & 13 00 56.6 & +37 48 16  & 16.90 & F  & H$_{\gamma}$, H$_{\beta}$, MgIb, H$_{\alpha}$ \\
\qq &HS 1408+3957           & 14 08 35.3 & +39 57 29  & 18.70 & F? & H$_{\beta}$, MgIb, H$_{\alpha}$  \\
\hline \\[-0.2cm]
\multicolumn{7}{l}{ $^\dagger$ objects from the second priority random sample; $^+$ APM magnitude} \\
\end{tabular}
\end{center}
\end{table*}




\scriptsize
\setcounter{qub}{0}

\begin{table*}[h]

\begin{center}
\caption{\label{Tab3a} Spectroscopic classification for objects from the
``random selected'' sample observed with the Calar Alto 2.2m telescope}

\begin{tabular}{rlll} \hline \\[-0.35cm]
\multicolumn{1}{c}{\#}               &
\multicolumn{1}{c}{Name}             &
\multicolumn{1}{r}{m$_{B}$}          &
\multicolumn{1}{c}{Type} \\
&
\multicolumn{1}{c}{ (1) } &
\multicolumn{1}{c}{ (2) } &
\multicolumn{1}{c}{ (3) } \\
\hline
\qq& HS 1007+3948            &17.40  & NOEM           \\
\qq& HS 1013+3804            &17.70  & Star          \\
\qq& HS 1029+4020            &16.70  & Star          \\
\qq& HS 1322+3708            &16.40  & BCG?          \\
\qq& HS 1355+3911            &19.20  & Star          \\
\qq& HS 1408+3524            &16.80  & ABS           \\
\qq& HS 1417+3921            &18.30  & NOEM           \\
\qq& HS 1420+3726            &17.50  & BCG?          \\
\qq& HS 1423+3945            &18.50  & SA in Sp      \\
\qq& HS 1459+3608            &18.80  & NOEM           \\
\qq& HS 1537+3520            &17.60  & NOEM           \\
\qq& HS 1545+3531            &18.00  & NOEM           \\
\qq& HS 1547+4008            &17.30  & NOEM           \\
\qq& HS 1601+3859            &18.40  & NOEM           \\
\qq& HS 1616+3627            &16.50  & Sy1           \\
\qq& HS 1649+3621            &18.10  & BCG?          \\
\qq& HS 1720+3929            &17.80  & NOEM           \\
\hline \\[-0.2cm]
\end{tabular}
\end{center}
\end{table*}





\end{document}